\documentclass[english]{article}
\usepackage[T1]{fontenc}
\usepackage[utf8]{inputenc}
\usepackage{amsmath}
\usepackage{amssymb}
\usepackage{babel}
\begin{document}
\title{A remark on the quasilocal calculation of tidal heating: energy transfer
through the quasilocal surface}
\author{Albert Huber\thanks{hubera@technikum-wien.at}}
\date{{\footnotesize{}UAS Technikum Wien - Department of Applied Mathematics
and Physics, H\"ochst\"adtplatz 6, 1200 Vienna, Austria}}
\maketitle
\begin{abstract}
In this note, using the quasilocal formalism of Brown and York, the
flow of energy through a closed surface containing a gravitating physical
system is calculated in a way that augments earlier results on the
subject by Booth and Creighton. To this end, by performing a variation
of the total gravitational Hamiltonian (bulk plus boundary part),
it is shown that associated tidal heating and deformation effects
generally are larger than expected. This is because the aforementioned
variation leads to previously unrecognized correction terms, including
a bulk-to-boundary inflow term that does not appear in the original
calculation of the time derivative of the Brown-York energy and leads
to corrective extensions of Einstein's quadrupole formula in the large
sphere limit. 
\end{abstract}
\textit{\footnotesize{}Key words: quasilocal energy, gravitational
Hamiltonian, tidal heating, bulk-to-boundary inflow}{\footnotesize\par}

\section*{Introduction}

The influence of tidal deformation and heating effects on a nearly
isolated gravitating systems due to external fields was successfully
treated and described by Booth and Creighton in \cite{booth2000quasilocal}.
Instead of using pseudotensorial techniques, which were more popular
at the time \cite{favata2001energy,purdue1999gauge,thorne1998tidal},
the authors used the Brown-York quasilocal formalism to address the
problem. Specifically, in order to describe the interaction of a spatially
and temporally bounded gravitating system with an external tidal field,
they performed a variation of the boundary part of the total gravitational
Hamiltonian, which allowed them to calculate energy and momentum fluxes
on a quasilocal surface and thus determine the tidal work done by
an external gravitational field in the so-called 'buffer zone'. Taking
advantage of the fact that the associated quasilocal energy expression
coincides with the Arnowitt-Deser-Misner energy at spatial infinity
\cite{arnowitt2008republication,brown1993quasilocal,mann2006covariant}
and the Trautman-Bondi-Sachs energy at null infinity \cite{bondi1962gravitational,brown1997energy,sachs1962gravitational,trautman1962gravitation},
the authors provided two interesting applications of their formulae. 

First, the authors calculated the energy flux due to gravitational
waves through a quasilocal surface in the wave zone (near infinity)
to obtain the correct mass loss for a system radiating gravitational
waves. 

Second, the authors calculated the tidal heating of a self-gravitating
body interacting with an external tidal field in the local asymptotic
reference frame of the body. While this effect typically occurs for
arbitrary pairs of bodies moving around each other in non-circular
orbits, the example discussed in \cite{booth2000quasilocal} is one
that is familiar from the solar system, namely the tidal deformation
and associated tidal heating of the Gallilean moon Io by Jupiter,
which is generally held responsible for the strong volcanic activity
on Jupiter's satellite. 

As will be shown in the second section of this work (after a brief
overview of the essentials of the considered quasilocal geometric
framework in the first section), particularly with respect to the
latter application, the analysis of Booth and Creighton can be extended
in one particular respect: The mass-energy transfer through the quasilocal
surface can alternatively be calculated by varying the total gravitational
Hamiltonian (bulk plus boundary parts), which, in contrast to the
results originally obtained by the authors, leads to the emergence
of correction terms, that is, a confined stress-energy term and a
bulk-to-boundary inflow term that combines the dynamical degrees of
freedom of the bulk with those of the boundary. Although this does
not change the validity of the original results of Booth and Creighton
(the authors considered a case in which the correction terms become
zero), it is nevertheless argued in the present work that especially
the latter term must be taken into account in the quasilocal calculation
of more violent tidal deformations and heating effects than those
treated in \cite{booth2000quasilocal}, especially in cases of tidal
deformations of celestial bodies revolving in circular motion around
each other, where there is considerable mass and/or radiation transfer
through the quasilocal surface (possibly due to the collapse of one
of the bodies). To illustrate this, a recap of Booth's and Creighton's
results on the tidal heating of isolated bodies is given in the third
section of this work, followed by a discussion of the extent to which
these results are insufficient to describe fluctuations of the interior
field of tidally deformed bodies that lead to nonnegligible backreactions
to the exterior fields of these objects. In this context, specific
ideas from the theory of extended irreversible thermodynamics are
used as the basis for the discussion. Apart from that, another comparatively
simple geometrical example is discussed, namely the mass-energy inflow
through the quasilocal surface of a spatially and temporally bounded
gravitating system, caused by gravitoelectromagnetic (GEM) fields.
Using this example, it is argued that corrections to the time derivative
of the Brown-York mass occur and lead to corrections to the Einstein
quadrupole formula in the large sphere limit; corrections that should
find application in Einstein-Hilbert gravity not only in the special
cases discussed in this paper, but in many cases of interest, including
the description of, for example, tidal deformation and tidal heating
effects caused by accretion phenomena or merger processes in relativistic
N-body systems.

\section{Geometric Setting and quasilocal Action}

As a prerequisite for later considerations, this section provides
a brief introduction to the quasilocal framework discussed in \cite{booth1999moving,booth2000quasilocal}
and review some of the important aspects of the underlying geometric
model. To that end, a spacetime $(\mathcal{M},g)$ with manifold $\mathcal{M}=M\cup\partial M$
shall be considered, which is foliated by a family of $t=const.$-hypersurfaces
$\{\Sigma_{t}\}$. Evolving continuously between two fixed instants
of time $\Sigma_{1}$ and $\Sigma_{2}$, this spacetime shall be assumed
to be temporally and spatially bounded. That is to say, it shall be
assumed to be bounded by a timelike hypersurface $\mathcal{B}$ in
such way that there holds $\partial M=\Sigma_{1}\cup\Sigma_{2}\cup\mathcal{B}$,
where none of the embedded boundary parts shall strictly be assumed
to be smooth; however, the timelike portion $\mathcal{B}$ shall be
assumed to be connected for the sake of simplicity. Moreover, the
timelike boundary part shall be assumed to be foliated by a collection
of two-surfaces $\{\Omega_{t}\}$ such that $\mathcal{B}=\{\underset{t}{\cup}\Omega_{t}:\,t_{1}\leq t\leq t_{2}\}$.
As a result of the fixings made, it is ensured that both spatial boundary
parts $\Sigma_{1}$ and $\Sigma_{2}$ are given in such a way that
they contain a pair of two-surfaces $\Omega_{1}$ and $\Omega_{2}$,
respectively, which bifurcate $\Sigma_{1}$ and $\Sigma_{2}$ locally.

The worldtube thus obtained divides spacetime into two distinct parts,
an interior part $\mathcal{M}$ and an exterior part $\bar{\mathcal{M}}$;
and that in such a way that the interior part $\mathcal{M}$ is contained
in the Lorentzian manifold $\bar{\mathcal{M}}$ of an associated ambient
spacetime $(\text{\ensuremath{\mathcal{\bar{M}}}},\bar{g})$ such
that $\mathcal{\mathcal{M}}\subset\bar{\mathcal{M}}$ and hence $\Sigma_{1},\Sigma_{2},\mathcal{B}\subset\bar{\mathcal{M}}$. 

Given this specific geometric setting, one can now select a time evolution
vector field $t^{a}$ which can be decomposed in components perpendicular
and parallel to the leaves of the spacelike foliation of the local
spacetime $(\mathcal{M},g)$. This means that $t^{a}$ can be written
- everywhere except at the timelike boundary $\mathcal{B}$ - in the
form $t^{a}=Nn^{a}+N^{a}$, where $N$ is the lapse function, $N^{a}$
is the shift vector field and $n^{a}$ represents a forward-pointing
timelike unit vector field orthogonal to the folia of $(\mathcal{M},g)$.
At the timelike boundary $\mathcal{B}$, however, the same vector
field must be decomposed alternatively in the form $t^{a}=\mathcal{N}v^{a}+\mathcal{N}^{a}$,
where $\mathcal{N}$ is the boundary lapse function, $\mathcal{N}^{a}$
is the boundary shift vector field and $v^{a}$ is a forward-pointing
timelike unit vector field being orthogonal to $\Omega_{t}$ and tangent
to $\mathcal{B}$. By the selection of a time flow vector field, a
time parameter $t$ can now be fixed by solving the relation $L_{t}t=1$
(according to which $L_{t}$ represents the Lie derivative along $t^{a}$). 

Next, it may be assumed that there exists an outward-pointing spacelike
unit normal to $\mathcal{B}$, henceforth denoted by $u^{a}$, which
is usually non-orthogonal to the timelike generator $n^{a}$, but
orthogonal to another timelike unit vector field $v^{a}$ that is
tangent to $\mathcal{B}$. At the same time, there is a spacelike
vector field $s^{a}$ which is orthogonal to the timelike generator
$n^{a}$. Due to the non-orthogonality of the two vector fields $n^{a}$
and $u^{a}$, there must exist a scalar field $\eta\equiv u_{a}n^{a}$,
which vanishes only for the special case of a timelike boundary which
is exactly orthogonal to all leaves of its spacelike foliation. This
scalar field also occurs in the relation $s_{a}v^{a}\equiv-\eta$
and in the decomposition relations $v^{a}=\lambda(n^{a}-\eta u^{a})$,
$s^{a}=\lambda(u^{a}+\eta n^{a})$ and $n^{a}=\lambda(v^{a}+\eta s^{a})$,
$u^{a}=\lambda(s^{a}-\eta v^{a})$, whose validity follows directly
from that of the normalization conditions $u^{a}u_{a}=s^{a}s_{a}=-n^{a}n_{a}=-v^{a}v_{a}=1$,
where $\lambda=\frac{1}{\sqrt{1+\eta^{2}}}$ is a boost parameter.
This boost parameter also occurs in the relation $\mathcal{N}=\lambda N$
between the standard lapse function $N$ and the boundary lapse function
$\mathcal{\mathcal{N}}$, which occur in the volume elements of spacetime
and that of the embedded spacelike boundary hypersurface, i.e. $\sqrt{-g}=N\sqrt{h}$
and $\sqrt{-\gamma}=\mathcal{N}\sqrt{q}$. These relations, in turn,
apply with respect to the determinant of the four-metric $g_{ab}$
of $(\mathcal{M},g)$, its induced three-metrics $h_{ab}=g_{ab}+n_{a}n_{b}$
on $\Sigma_{t}$ and $\gamma_{ab}=g_{ab}-u_{a}u_{b}$ on $\mathcal{B}$
and the two-metric $q_{ab}=g_{ab}+n_{a}n_{b}-s_{a}s_{b}=g_{ab}-u_{a}u_{b}+v_{a}v_{b}$
that is induced by $g_{ab}$ on $\Omega_{t}$. 

Ultimately, as a byproduct of considering all these definitions, Hayward's
quasilocal action \cite{hayward1993gravitational}

\begin{equation}
S_{G}[g]=\frac{1}{16\pi}(\underset{M}{\int}R\omega_{g}+\underset{\Sigma_{1}}{2\overset{\Sigma_{2}}{\int}}K\omega_{h}+2\underset{\mathcal{B}}{\int}\Theta\omega_{\gamma}+\underset{\Omega_{1}}{2\overset{\Omega_{2}}{\int}}\sinh^{-1}\eta\omega_{q})
\end{equation}
can finally be set up, where, in this context, $R=g^{ab}R_{ab}$ denotes
the Ricci scalar of the spacetime $(\mathcal{M},g)$ and $K=h^{ab}K_{ab}$
and $\Theta=\gamma^{ab}\Theta_{ab}$ term the scalar extrinsic curvatures
of the spacelike folia $\Sigma_{t}$ and the timelike boundary $\mathcal{B}$
of $(\mathcal{M},g)$. Moreover, as can be seen, the different volume
elements $\omega_{g}:=\sqrt{-g}d^{4}x$, $\omega_{h}:=\sqrt{h}d^{3}x$,
$\omega_{\gamma}:=\sqrt{-\gamma}dtd^{2}x$ and $\omega_{q}:=\sqrt{q}d^{2}x$
enter the definition of this quasilocal action.

Following closely the steps taken in \cite{booth1999moving,brown1993quasilocal},
one finds that Hayward's action can be re-written in a slightly different
form. More specfically, using the decomposition relation $R=^{(3)}R+K_{ab}K^{ab}-K^{2}+2\nabla_{b}(Kn^{b}-a^{b})$
for the Ricci scalar, the the Hamiltonian and momentum constraints
$\mathcal{H}=\frac{8\pi}{\sqrt{h}}(P_{ab}P^{ab}-\frac{1}{2}P^{2})-\frac{\sqrt{h}}{16\pi}{}^{(3)}R$
and $\mathcal{H}{}_{a}=-2D_{b}P_{\;a}^{b}$, which are defined with
respect to the fields $P^{ab}=\frac{\sqrt{h}}{16\pi}(K^{ab}-h^{ab}K)$,
$K_{ab}=\frac{1}{2N}(\dot{h}_{ab}-2D_{(a}N_{b)})$ and $\frac{\sqrt{-g}}{16\pi}(K_{ab}K^{ab}-K^{2})=P_{ab}\dot{h}^{ab}-\frac{8\pi N}{\sqrt{h}}(P_{ab}P^{ab}-\frac{1}{2}P^{2})-2P_{ab}D^{a}N^{b}$,
as well as the identities $k=\lambda(\Theta+\eta K-a_{b}u^{b}+\lambda(v\nabla)\eta)$,
$Nk-(K_{ab}-Kh_{ab})N^{a}s^{b}=\mathcal{N}\mathfrak{K}-\lambda(\mathcal{N}\nabla)\eta-(\mathcal{N}\nabla)v_{a}u^{a}$,
in relation to which $\mathfrak{K}=q^{ab}\mathfrak{K}_{ab}=q^{ab}\nabla_{a}u_{b}$
represents the extrinsic curvature of the spacelike intersection surfaces
$\Omega_{t}$ formed in relation to $u^{a}$, one finds that Hayward's
action can be rewritten in the form

\begin{align}
S_{G}[g] & =\int dt\underset{\Sigma_{t}}{\int}d^{3}x[P_{ab}h^{ab}-N\mathcal{H}-\mathcal{H}_{a}N^{a}]+\\
 & +\frac{1}{8\pi}\int dt\underset{\Omega_{t}}{\int}\dot{\omega}_{q}\sinh^{-1}\eta+\int dt\underset{\Omega_{t}}{\int}d^{2}x[\mathcal{N}\mathfrak{h}+\mathfrak{h}_{a}\mathcal{N}^{a}],\nonumber 
\end{align}
where $\mathfrak{h}=\frac{\sqrt{q}}{8\pi}\mathfrak{K}$, $\mathfrak{h}_{a}=-\frac{\sqrt{q}}{8\pi}u^{c}q_{\;a}^{b}\nabla_{b}v_{c}$
and the overdot denotes the Lie derivative with respect to $t^{a}$. 

As a result of these conversions of Hayward's quasilocal action, the
generalized gravitational Hamiltonian

\begin{equation}
H=\underset{\Sigma_{t}}{\int}(N\mathcal{H}+\mathcal{H}_{a}N^{a})d^{3}x-\underset{\Omega_{t}}{\int}(\mathcal{N}\mathfrak{h}+\mathfrak{h}_{a}\mathcal{N}^{a})d^{2}x
\end{equation}
is obtained, whose precise form was first determined by Brown and
York in the course of their field-theoretical generalization of Hamilton-Jacobi
theory \cite{brown1993quasilocal}. Given with respect to the canonical
variables $(h^{ab},P_{ab})$ and $(\sqrt{q},\sinh^{-1}\eta)$, this
generalized gravitational Hamiltonian provides a feasible quasilocal
description of the total amount of energy-momentum of a gravitational
field with spatially and temporally finite dimensions, which represents
an alternative to more traditional pseudotensorial descriptions of
the subject\footnote{For a detailed review, see for instance \cite{szabados2004quasi}.}. 

An important characteristic of the given generalized Hamiltonian is
the fact that it splits up into a bulk part $H_{Bulk}=\underset{\Sigma_{t}}{\int}(N\mathcal{H}+\mathcal{H}_{a}N^{a})d^{3}x$
and a boundary part $H_{Boundary}=-\underset{\Omega_{t}}{\int}(\mathcal{N}\mathfrak{h}+\mathfrak{h}_{a}\mathcal{N}^{a})d^{2}x$,
where, as already indicated, the boundary part $H_{Boundary}$ coincides
in an asymptotically flat spacetime in the so-called large sphere
limit with well-known candidates for the total quasilocal stress-energy-momentum
of the gravitational field itself - with that of Arnowitt, Deser and
Misner at spacelike infinity and that of Bondi and Sachs at null infinity.

Given the exact form of the quasilocal gravitational Hamiltonian $(3),$
the quasilocal charge 
\begin{equation}
E=\underset{\Sigma_{t}}{\int}(N\varepsilon-p_{a}N^{a})\omega_{h}-\underset{\Omega_{t}}{\int}(\mathcal{N}e-j_{a}\mathcal{N}^{a})\omega_{q}
\end{equation}
can be derived by using Einstein's equations 
\begin{equation}
G_{ab}=8\pi T_{ab}.
\end{equation}
As may be noted, the quasilocal charge obtained again consists of
two parts; a bulk part $E_{Bulk}=\underset{\Sigma_{t}}{\int}(N\varepsilon-p_{a}N^{a})\omega_{h}$
and a boundary part $E_{Boundary}=-\underset{\Omega_{t}}{\int}(\mathcal{N}e-j_{a}\mathcal{N}^{a})\omega_{q}$.

This all is of significance in that the results obtained so far form
the starting point for Booth and Creighton's quasilocal calculation
of tidal heating, which will be briefly discussed in the third and
final section of this paper. However, for the analysis of the results
of the now upcoming second section of this work, it is important to
note that the autors considered only the boundary part of expression
$(4)$ as the basis for their quasilocal calculation of tidal heating.
More specifically, an important step in their calculation was to determine
the temporal variation of said boundary part, for which they obtained
the result
\begin{equation}
\frac{dW}{dt}=\underset{\Omega_{t}}{\int}(-\dot{\mathcal{N}}e+j_{a}\dot{\mathcal{N}}^{a}+\frac{\mathcal{N}}{2}s^{ab}\dot{q}_{ab})\omega_{q}.
\end{equation}
As will be shown in the next section, the calculation yields a different
result when the full quasilocal expression $(4)$ (and only its boundary
part) is considered. Specifically, it is shown that the temporal variation
of the quasilocal charge $(4)$ leads to correction terms that do
not appear in the original calculation of Booth and Creighton, but
must typically be included when considering non-vanishing stress-energy
tensors. As is argued later in the third and final section of this
work, this may prove to be of relevance not only for the determination
of tidal heating effects, but also for other applications in the quasilocal
approach to general relativity.

\section{Gravitational Hamiltonian and quasilocal Energy Flux }

With the results of the previous section at hand, the next step will
be to derive the exact form of the correction terms mentioned in the
introduction. For this purpose, a temporal variation of the total
functional $E$ (bulk plus boundary) shall be calculated in the following
by Lie dragging it along the flow of the time evolution vector field
$t^{a}$. As a basis for doing this, the decomposition relations $T_{ab}=\varepsilon n_{a}n_{b}+n_{a}p_{b}+p_{a}n_{b}+S_{ab}$
and $\tau_{ab}=ev_{a}v_{b}+v_{a}j_{b}+j_{a}v_{b}+s_{ab}$ will be
used, whose validity implies that $T_{ab}t^{a}n^{b}=N\varepsilon-p_{a}N^{a}$
and $\tau_{ab}t^{a}v^{b}=\mathcal{N}e-j_{a}\mathcal{N}^{a}$ holds
true with respect to any given pair of (totally general) stress-energy
and surface stress-energy tensors $T_{ab}$ and $\tau_{ab}$. As may
be noted in this context, the fields $p_{a}=h_{a}^{\:c}T_{cb}n^{b}$
and $j_{a}=q_{a}^{\:c}\tau_{cb}v^{b}$ represent the mass-energy and
surface mass-energy fluxes and $S_{ab}=h_{a}^{\:c}h_{b}^{\:d}T_{cd}$
and $s_{ab}=q_{a}^{\:c}q_{b}^{\:d}\tau_{cd}$ represent the stress
and surface stress tensors of matter fields located within the bounded
spacetime region $\mathcal{M}\subset\mathcal{\bar{M}}$.

Assuming that $\mathsf{\mathfrak{L}}_{t}$ denotes the induced Lie-derivative
at $\Sigma_{t}$ pointing along $t^{a}$, it proves to be instrumental
for the calculation of the time derivative of $E$ to note that
\begin{align}
\mathsf{\mathfrak{L}}_{t}[\omega_{h}(N\varepsilon-p_{a}N^{a})] & =\omega_{h}\mathsf{\mathfrak{L}}_{t}(N\varepsilon-p_{a}N^{a})+\nonumber \\
 & +\omega_{h}(NK+D_{b}N^{b})(N\varepsilon-p_{a}N^{a})
\end{align}
holds true with respect to any spacetime $(\mathcal{M},g)$ whose
geometric structure permits consideration of the quasilocal approach
described in the very first section. Moreover, it proves beneficial
to heed another observation: The local conservation laws $\nabla_{a}T_{\;c}^{a}\cdot Nn^{c}=\nabla_{a}T_{\;c}^{a}\cdot N^{c}=0$,
in combination with the decompositions of $T_{ab}$ and $\tau_{ab}$
mentioned above, lead to the generalized continuity and contracted
generalized Euler equations
\begin{equation}
\mathsf{\mathfrak{L}}_{t}\varepsilon=(ND)\varepsilon-\varepsilon NK-ND_{a}p^{a}-2(pD)N-NS_{ab}K^{ab}
\end{equation}
and

\begin{equation}
\mathsf{\mathfrak{L}}_{t}p_{a}\cdot N^{a}=(ND)(p_{a}N^{a})-ND_{b}S_{\;a}^{b}N^{a}-S_{\;a}^{b}N^{a}D_{b}N-NKp_{a}N^{a}-\varepsilon(ND)N;
\end{equation}
equations that were first published (though in terms of some slightly
different conventions) many years ago in \cite{york1979kinematics}.
For notational clarification, it may be noted here that the definitions
$(ND):=N^{a}D_{a}$ and $(pD):=p^{a}D_{a}$ have been used in the
present context. 

By suitably combining these relations and applying the inverse Leibniz
rule with the idea of producing a total divergence term, one obtains
the result
\begin{align}
N\mathsf{\mathfrak{L}}_{t}\varepsilon-\mathsf{\mathfrak{L}}_{t}p_{a}\cdot N^{a} & =D_{a}[(N\varepsilon-p_{b}N^{b})N^{a}-N^{2}p^{a}+NS_{\;b}^{a}N^{b}]-\\
 & -(NK+D_{a}N^{a})(N\varepsilon-p_{b}N^{b})-NS_{ab}(NK^{ab}+D^{(a}N^{b)}).\nonumber 
\end{align}
Therefore, in further consequence, one finds 
\begin{align}
\mathsf{\mathfrak{L}}_{t}[\omega_{h}(N\varepsilon-p_{a}N^{a})] & =\omega_{h}(\mathsf{\mathfrak{L}}_{t}N\varepsilon-p_{a}\mathsf{\mathfrak{L}}_{t}N^{a}+\frac{N}{2}S_{ab}\mathsf{\mathfrak{L}}_{t}h^{ab})+\\
 & +\omega_{h}D_{a}[(N\varepsilon-p_{b}N^{b})N^{a}-N^{2}p^{a}+NS_{\;b}^{a}N^{b}].\nonumber 
\end{align}
By applying Gauss' law in the next step and using the constraint relation
$\mathcal{D}_{a}\tau_{\:b}^{a}=\pi^{a}=\gamma_{\:b}^{a}T_{ac}u^{c}$,
in respect to which $\mathcal{D}_{a}\tau_{\:b}^{a}=\gamma_{\:b}^{d}\gamma_{\:c}^{e}\nabla_{e}\tau_{\,d}^{c}$
applies, the temporal variation of the functional $E$ can now straightforwardly
be calculated, leading to the result
\begin{equation}
\frac{dE}{dt}=\underset{\Sigma_{t}}{\int}(\dot{N}\varepsilon-p_{a}\dot{N}^{a}+\frac{N}{2}S_{ab}\dot{h}^{ab})\omega_{h}-\underset{\Omega_{t}}{\int}(\dot{\mathcal{N}}e-j_{a}\dot{\mathcal{N}}^{a}+\frac{\mathcal{N}}{2}s_{ab}\dot{q}^{ab}-\mathfrak{X})\omega_{q}.
\end{equation}
Here, it should be noted that to obtain this very integral expression,
the definitions $s_{ab}=q_{a}^{\:c}q_{b}^{\:d}\tau_{cd}$, $b^{a}\equiv(v\nabla)v^{a}$,
$\dot{N}=\mathsf{\mathfrak{L}}_{t}N$, $\dot{N}^{a}=\mathsf{\mathfrak{L}}_{t}N^{a}$,
$\dot{h}^{ab}=\mathsf{\mathfrak{L}}_{t}h^{ab}$ as well as $\dot{\mathcal{N}}=\mathcal{L}_{t}\mathcal{N}$,
$\dot{\mathcal{N}}^{a}=\mathcal{L}_{t}\mathcal{N}^{a}$, $\dot{q}^{ab}=\mathcal{L}_{t}q^{ab}$
have been used, where $\mathcal{L}_{t}$ denotes the induced Lie-derivative
at $\Omega_{t}$ pointing along $t^{a}$. 

The integrand $\mathfrak{X}$ occurring in the boundary part of expression
$(12)$ consists of two parts, i.e. 
\begin{equation}
\mathfrak{X}=\mathfrak{X}_{1}+\mathfrak{X}_{2},
\end{equation}
where 
\begin{equation}
\mathfrak{X}_{1}=\mathcal{N}^{2}\pi_{a}v^{a}+\mathcal{N}\pi_{a}\mathcal{N}^{a}
\end{equation}
represents the confined stress-energy term and 
\begin{equation}
\mathfrak{X}_{2}=-\frac{\eta}{\lambda}\mathcal{N}^{2}\varepsilon+\eta\mathcal{N}^{2}p_{a}v^{a}+\eta\mathcal{N}p_{a}\mathcal{N}^{a}-\frac{\mathcal{N}^{2}}{\lambda}p_{a}u^{a}+\lambda\mathcal{N}S_{ab}v^{a}u^{b}+\lambda S_{ab}\mathcal{N}^{a}u^{b}
\end{equation}
represents a bulk-to-boundary inflow term, as may be concluded from
the fact that the bulk expressions $\varepsilon$, $p_{a}$ and $S_{ab}$
enter into its definition\footnote{Note that the bulk fields $\varepsilon$, $p_{a}$ and $S_{ab}$ can,
of course, be re-expressed on the boundary in terms of associated
boundary fields. However, in order to highlight the fact that these
fields can be used to interpret $\underset{\Omega_{t}}{\int}\mathfrak{X}_{2}\omega_{q}$
as a bulk-to-boundary inflow, the form of $\varepsilon$, $p_{a}$
and $S_{ab}$ has been intentionally left unchanged at this point.}. For the sake of clarification, it may be noted here that the relations
$N_{a}s^{a}=-\eta\mathcal{N}$, $N\varepsilon-p_{a}N^{a}=\frac{\mathcal{N}}{\lambda}\varepsilon-\mathcal{N}p_{a}v^{a}-p_{a}\mathcal{N}^{a}$,
$N^{2}p_{a}s^{a}=\frac{\mathcal{N}^{2}}{\lambda}p_{a}u^{a}$ and $S_{ab}s^{a}N^{b}=\lambda\mathcal{N}S_{ab}v^{a}u^{b}+\lambda S_{ab}\mathcal{N}^{a}u^{b}$
have been used in order to obtain this particular form of equation
$(15)$. The corresponding integral term that occurs in relation $(12)$,
which results from converting the total divergence appearing in equation
$(11)$, is equal to the net flux of quasilocal bulk energy passing
through $\Omega_{t}$ being transferred from the gravitating physical
system under consideration (through the spatial boundary) to its environment. 

Upon closer inspection of the full boundary part of integral expression
$(12)$, it can further be observed that said part of the expression
(on account of its physical units) can be interpreted as a power functional
of the form $\mathcal{P}=\underset{\Omega_{t}}{\int}\mathcal{I}\omega_{q}$,
where the corresponding integrand reads

\begin{equation}
\mathcal{I}=\dot{\mathcal{N}}e-j_{a}\dot{\mathcal{N}}^{a}+\frac{\mathcal{N}}{2}s_{ab}\dot{q}^{ab}-\mathfrak{X}.
\end{equation}
Sure enough, however, such an interpretation only truly makes sense
if electromagnetic and/or gravitational radiation passes through the
quasilocal surface $\Omega_{t}$; a case in which the quantity $\mathcal{I}$
typically can be identified as the intensity of the total radiant
energy escaping from the system into the environment. But as shall
be discussed in greater detail in the next section, where applications
will be considered, the integral $\underset{\Omega_{t}}{\int}\mathfrak{X}_{2}\omega_{q}$
cannot be neglected.

Before that, however, equation $(12)$ shall first be considered from
a different angle in order to also illuminate the geometric aspect
of the chosen approach. For this purpose, one may rewrite $(12)$
using the field equations of the theory, which yields the result

\begin{align}
\frac{dH}{dt} & =\underset{\Sigma_{t}}{\int}(\dot{N}\mathcal{H}-\mathcal{H}_{a}\dot{N}^{a}+\frac{N}{2}\mathcal{Q}_{ab}\dot{h}^{ab})d^{3}x-\\
 & -\underset{\Omega_{t}}{\int}(\dot{\mathcal{N}}\mathfrak{h}-\mathfrak{h}_{a}\dot{\mathcal{N}}^{a}+\frac{\mathcal{N}}{2}\mathfrak{Q}_{ab}\dot{q}^{ab}-\sqrt{q}\mathfrak{X})d^{2}x\nonumber 
\end{align}
for the temporal variation of the gravitational Hamiltonian. As may
be noted, the definitions
\begin{align}
\mathcal{Q}_{ab} & =\frac{\sqrt{h}}{8\pi}\{{}^{(3)}R_{ab}-2K_{ac}K_{\;b}^{c}+KK_{ab}-\frac{1}{N}\left[\dot{K}_{ab}+(ND)K_{ab}+D_{a}D_{b}N\right]-\nonumber \\
 & -\frac{1}{2}h_{ab}\left(^{(3)}R+K^{2}-K_{cd}K^{cd}-\frac{2}{N}\left[\dot{K}+(ND)K+D_{a}D^{a}N\right]\right)\}
\end{align}
and
\begin{equation}
\mathfrak{Q}_{ab}=\frac{\sqrt{q}}{8\pi}(\mathfrak{K}_{ab}-(\mathfrak{K}-b_{a}u^{a})q_{ab})
\end{equation}
have been used in the present context. Moreover, as may also be noted,
relation $(18)$ has been used in combination with the constraint
equations of the theory to rewrite relations $(13)$ and $(14)$ in
the form 
\begin{equation}
\mathfrak{X}_{1}=\mathcal{N}^{2}\mathcal{D}_{a}\tau_{\;b}^{a}v^{b}+\mathcal{N}\mathcal{D}_{a}\tau_{\;b}^{a}\mathcal{N}^{a}
\end{equation}
and
\begin{align}
\mathfrak{X}_{2} & =-\frac{\eta}{\lambda}\mathcal{N}^{2}\mathcal{H}-\frac{1}{2}\eta\mathcal{N}^{2}\mathcal{H}_{a}v^{a}-\frac{1}{2}\eta\mathcal{N}\mathcal{H}_{a}\mathcal{N}^{a}-\\
 & +\frac{\mathcal{N}^{2}}{2\lambda}\mathcal{H}_{a}u^{a}+\lambda\mathcal{N}\mathcal{Q}_{ab}v^{a}u^{b}+\lambda\mathcal{Q}_{ab}\mathcal{N}^{a}u^{b}\nonumber 
\end{align}
and therefore obtain with $(17)$ a purely geometrical expression
that agrees exactly with the time derivative of the quasilocal gravitational
Hamiltonian. Here, as may be noted, the use of Einstein's equations
(or rather the constraint equations resulting from them) does not
prove to be mandatory in this context, but rather practical, since
all calculations could also be carried out exclusively with geometric
quantities, which, of course, yields exactly the same result in the
end. 

As may be noted, there is something both astonishing and appealing
about results $(12)$ and $(17)$, namely, that the occurring bulk
and boundary integral expressions look very similar; only the correction
terms prevent an exact similarity. But not only this similarity is
interesting: From the structure of $(17)$ it can be generally concluded
that a change of the matter content of a spatially and temporally
bounded gravitating system always depends on how the individual components
of the metric evolve over time. Moreover, it can be concluded that
such a change also strongly depends on the structure of the occurring
correction terms, which can be interpreted physically as a bounded
mass energy distribution plus a non-vanishing bulk to boundary inflow
that combines the dynamical degrees of freedom of the bulk with those
of the boundary; even if it is assumed that the considered local gravitational
field is stationary. It can be expected that pseudotensorial methods
lead to similar results, although further studies of such an approach
may be in order. 

Anyway, to illustrate the physical relevance of the derivations made,
some selected applications will now be discussed in the next section,
with a particular focus on possibly testable implications of the theory.
In particular, two effects will be examined in more detail, namely
the loss of quasilocal mass energy due to $i)$ tidal deformation
and heating, and $ii)$ the emission of gravitational, electromagnetic,
and/or gravitoelectromagnetic radiation. With respect to the first
type of effect, as it turns out, the model proves consistent with
the results of Booth and Creighton in the sense that the derived correction
terms play no role in calculating the loss of quasilocal mass due
to tidal heating. The reason for this, as will be shown, is that the
authors' approach remains fully valid precisely when the source-free
Einstein equations are considered (as done in \cite{booth2000quasilocal}),
whereas it must very well be extended if matter fields are present
and thus the non-vacuum field equations apply. However, this is exactly
what is to be expected in the presence of tidal effects which lead
to perturbations not only of the external but also of the internal
field of a body. Something similar is also to be expected, as shall
be made clear below, in the second case to be discussed, which is
devoted to the combined emission of gravitational and electromagnetic
radiation by a gravitating physical system.

\section{Tidal Heating, Gravitoelectromagnetism and further Applications }

In view of the results derived in the previous section, the question
naturally arises whether or not the derived correction terms play
a role for the quasilocal calculation of tidal heating, and whether
or not, therefore, the original results of Booth and Creighton from
\cite{booth2000quasilocal} need to be modified or generalized in
some sense. Unsurprisingly, as will be explained below, the authors'
approach turns out to be perfectly sound (in terms of the specific
assumptions made in their work). However, as shall also be made clear
below, said approach needs very well to be generalized in cases where
quasilocal methods are to be used to describe more violent tidal deformations
and heating effects than those discussed in \cite{booth2000quasilocal},
such as in cases of tidal deformations of celestial bodies in $N$-body
systems that are accompanied by substantial mass and/or radiation
transfer through the quasilocal surface, as occur for example when
one of the bodies collapses or pairs of bodies merge with each other.

Before discussing mass loss due to such effects, however, it is first
necessary to show that Booth's and Creighton's results occur as a
special case of those derived in the previous section. To show this,
it is sufficient to make a single assumption, namely that the considered
spacetime $(\mathcal{M},g)$ is Ricci-flat and therefore the on-shell
vacuum field equations

\begin{equation}
R_{ab}=0
\end{equation}
are satisfied; a requirement that has the effect that the time derivative
of the bulk part appearing in $(12)$ becomes zero in the same way
as the derived correction terms, which further implies that this relation
reduces to the form 

\begin{equation}
\frac{dE}{dt}=-\underset{\Omega_{t}}{\int}(\dot{\mathcal{N}}e-j_{a}\dot{\mathcal{N}}^{a}+\frac{\mathcal{N}}{2}s_{ab}\dot{q}^{ab})\omega_{q}.
\end{equation}
Consequently, however, by utilizing the fact that $\dot{q}^{ab}=-q^{ac}q^{bd}\dot{q}_{cd}$
holds in the given context, one finds that relations $(6)$, $(12)$
and $(23)$ all coincide in the case that the source-free Einstein
equations are satisfied, thus making it clear that the results of
the work of Booth and Creighton occur as a special case of the results
derived in section two of this work. Based on this, it can be further
concluded that the temporal variation of the full Hamiltonian (bulk
plus boundary) needs to be considered only when matter fields are
present and spacetime is not Ricci-flat, while in the opposite case
it is sufficient to consider the boundary part of the gravitational
Hamiltonian, as was done in \cite{booth2000quasilocal}. 

Having clarified this, it shall now be briefly be reviewed how the
obtained formula can be used to describe tidal deformation and heating
effects. To this end, one may recall that Booth and Creighton used
the Hartle-Thorne model \cite{hartle1968slowly,thorne1998tidal} as
a basis for their quasilocal calculations, i.e., a perturbative vacuum
model that describes how the external field of a small self-gravitating
central body changes under the influence of an external tidal field.
This very model deals with a spacetime metric $g_{ab}$ of the form

\begin{equation}
g_{ab}=\eta_{ab}+e_{ab},
\end{equation}
which is given with respect to the flat Minkowski metric $\eta_{ab}$,
where $\vert e_{ab}\vert\ll1$ applies by definition. In rectangular
coordinates, the individual components of the perturbation tensor
$e_{ab}$ take the form $e_{00}\equiv\phi\equiv-\frac{2M}{r}-\frac{3Q_{ik}s^{i}s^{k}}{r}+r^{2}\mathcal{E}_{ik}s^{i}s^{k}$,
$e_{0j}\equiv A_{j}\equiv-\frac{2}{r}\frac{dQ_{jk}}{dt}s^{k}+\frac{4r^{3}}{21}\frac{d\mathcal{E}_{jk}}{dt}s^{k}-\frac{10r^{3}}{21}\frac{d\mathcal{E}_{ik}}{dt}s^{i}s^{k}s_{j}$
and $e_{ij}=(\frac{2M}{r}-\frac{3}{r^{3}}Q_{kl}s^{k}s^{l}+r^{2}\mathcal{E}_{kl}s^{k}s^{l})\delta_{ij}$,
where, in this regard, the field $Q_{ik}=\int\varepsilon(x_{i}x_{k}-\frac{1}{3}r^{2}\delta_{ik})d^{3}x$
represents the tracefree and symmetric quadrupole moment of the self-gravitating
body and $\mathcal{E}_{ik}=\mathcal{R}_{i0k0}$ the electric part
of the Weyl tensor, which coincides with the Riemann curvature tensor
in the given case. The radial unit vector occurring in this context
is given by $s^{i}=\frac{x^{i}}{r}$, where $r$ is the radial distance
from the center of the isolated body as measured in its local asymptotic
rest frame. 

Taking advantage of the gauge ambiguity in setting up a nearly Minkowskian
coordinate system and switching to spherical coordinates, it was shown
in \cite{booth2000quasilocal} that the Hartle-Thorne metric $(24)$
can be converted into a slightly different form. More specifically,
by using the diffeomorphism freedom to make the replacement $e_{ab}\rightarrow e_{ab}+2\partial_{(a}\xi_{b)}$
with respect to the co-vector field $\xi_{j}=\alpha r^{-2}Q_{jk}s^{k}+\beta r^{3}\mathcal{E}_{jk}s^{k}+\gamma r^{3}\mathcal{E}_{ik}s^{i}s^{k}s_{j}$,
where $\alpha$, $\beta$ and $\gamma$ are free constants of order
one, it was shown it the aforementioned work that the rate at which
tidal work is performed on an isolated body is given (modulo higher
order terms) by the same formula as in the post-Newtonian approximation
of the theory, which had previously been derived by using pseudotensorial
techniques \cite{favata2001energy,purdue1999gauge}. Using here the
shorthand notation $\mathfrak{Y}=\frac{1}{60}\frac{d}{dt}\left[2ar^{5}\mathcal{E}_{ik}\mathcal{E}^{ik}+2b\mathcal{E}_{ik}Q^{ik}-cr^{5}Q_{ik}Q^{ik}\right]$
in conjunction with the definitions $a=8\beta\gamma+4\gamma^{2}-2\beta^{2}+4\gamma-2\beta-3,b=8\alpha\gamma-12\gamma+6\beta-2\alpha+3$
and $c=4\alpha^{2}-12\alpha-9$, the obtained formula reads

\begin{equation}
\frac{dE}{dt}=\frac{1}{2}\mathcal{E}_{ik}\frac{dQ^{ik}}{dt}+\mathfrak{Y}.
\end{equation}
In this context, the $\frac{1}{2}\mathcal{E}_{ik}\frac{dQ^{ik}}{dt}$-term
specifies the irreversible (unrecoverable) part of the work done to
deform and heat the system (the dissipated energy is converted into
heat), whereas the second $\mathfrak{Y}$-term represents the reversible
(recoverable) part of the work being done to increase the potential
energy of the system. Regarding these two terms, the former irreversible
one turns out to be the one of primary interest, since $i)$ it is
invariant under diffeomorphisms genererated by $\xi_{j}$ (in contrast
to the second reversible term) and $ii)$ it is the same leading order
term that was previously obtained through analogous calculation using
pseudotensorial instead of quasilocal methods (which, surprisingly,
all lead to the same expression regardless of the choice of energy-momentum
complex and gauge conditions \cite{favata2001energy,purdue1999gauge};
a situation previously known only from the case of pseudotensors calculated
from Kerr-Schild metrics \cite{leaute1976energy,Virbhadra:1991ba,Xulu:2007vg}).
Although not really obvious at the time, this compatibility of results
is not very surprising from today's viewpoint in that it has been
shown by Chang, Chen and Nester \cite{chang1999pseudotensors} that
any energy-momentum pseudotensor leads to an associated Hamiltonian
boundary term, so that it becomes clear that - within the quasilocal
Hamiltonian framework - pseudotensors either give rise to distinct
Dirichlet type boundary conditions or to Neumann type boundary conditions;
all of which are algebraic in terms of the metric. Consequently, however,
since the tidal heating results obtained in the literature appear
to be independent of the choice of boundary conditions, it is hardly
surprising that the purely quasilocal calculation of Booth and Creighton
leads to very similar results as those of earlier works on the subject
that used the pseudotensorial approach. This is not least the case
because, in the approximate calculations mentioned, the precise form
of the boundary conditions is not essential, because the results obtained
by using different boundary conditions deviate from each other only
in higher order. Therefore, from a purely mathematical point of view,
the good agreement of the results proves to be completely natural. 

However, what is striking in this context is that, in contrast to
the results of earlier pseudotensorial approaches, the quasilocal
results are not only completely general but also coordinate-independent,
which is why the results obtained could in principle also be used
to determine tidal deformations and heating effects in more sophisticated
and physically challenging situations; that is, in situations where
not only fluctuations of the external gravitational field of the body
are considered, but also changes of the interior field of the self-gravitating
body are taken into account. After all, it has to be kept in mind
that the Thorne-Hartle model is not designed to describe directly
the tidal deformations of a self-gravitating body, but only the backreaction
on its external gravitational field. Backreactions to the interior
field of the body are not accounted for by the model, which is why
it can only be used to determine the effects of tidal heating if the
overall structural changes of the body are sufficiently small and
thus negligible. 

Situations in which disregarding such structural changes no longer
seems justified are those in which the shape, composition, and macroscopic
thermal state of a given body (or collection of bodies) change drastically
during gravitational interaction, e.g., when the body becomes heavily
tidally deformed and heated during interaction with another, more
massive celestial body, which may eventually lead to the collapse
and disintegration of the body and its accretion by its much more
massive companion. In such a situation, with both celestial bodies
typically orbiting each other in circular motion, it is reasonable
to expect that matter is constantly escaping from the system through
the quasilocal surface of spacetime; at least as long as one assumes
(inspired by Newton's theory of gravity) that the quasilocal surface
represents the body's gravisphere, i.e., the spatially bounded sphere
of influence of the collapsing body under consideration. Certainly
also the collision of bodies - a process in which similarly strong
tidal deformations can occur - cannot be described on the basis of
a vacuum model like that of Hartle and Thorne. 

It must be acknowledged, however, that describing the aforementioned
phenomenological effects (even at the numerical level) is an extremely
complicated venture; especially given the need to describe how the
body's internal gravitational field changes such that the matter content
of the spatially and temporally bounded gravitational system changes
relative to its local physical environment. 

Yet, despite the lack of a geometric model that could fully account
for these effects, it shall be argued below that the integral expression
$(12)$ of the previous section can indeed be used to predict the
behavior of very general types of matter distributions whose general
form appears suitable to describe (at least in principle) processes
in which objects are very strongly tidally deformed due to the phenomenological
effects mentioned above. Moreover, it will be argued why and to what
extent the Hartle-Thorne approximation proves insufficient to describe
gravitational backreactions caused by isolated bodies that fail to
be in almost perfect (macroscopic) internal thermal equilibrium.

As a basis for the arguments used, the theory of extended irreversible
thermodynamics \cite{hiscock1983stability,hiscock1985generic,israel1976nonstationary,israel1979transient,jou1999extended,landau1987theoretical}
shall be used at this point, with the main idea being that the corresponding
physical framework is general enough to consider the case where tidal
heating of a body (or collection of bodies) leads to the occurrence
of heat flows and viscous stresses, and thus to non-negligible backreactions
on the internal geometric field of the body (or bodies). In the mentioned
theory, the matter distribution under consideration is that of a fluid
mixture. Such a matter distribution can be fully characterized by
three different quantities, i.e. the so-called thermal energy-momentum
tensor $T_{\;b}^{a}=\epsilon w^{a}w_{b}+\varpi^{a}w_{b}+w^{a}\varpi_{b}+(p+\Pi)h{}_{\,b}^{a}+\Pi{}_{\,b}^{a}$,
a particle number current $N_{A}^{a}=n_{A}w^{a}+\nu_{A}^{a}$, defined
with respect to a number $A$ of species of particles, and a so-called
covariant entropy current $S^{a}=sw^{a}+\eta^{a}$, where, in this
context, the scalar fields $\epsilon$, $p$ and $s$ represent the
energy density, pressure and total entropy density of the fluid mixture,
$n_{A}$ is the particle density of a number of particle species $A$
and $\pi$ represents the viscous bulk pressure. The vector fields
$w^{a}$, $\varpi^{a},$ $\nu_{A}^{a}$ and $\eta^{a}$ meanwhile
represent the four-velocity vector of the system, an energy flux current
and so-called particle diffusion and heat conduction fluxes, while
the tensor field $\pi{}_{\,b}^{a}$ represents the trace-free anistropic
viscous stress-tensor\footnote{As may be noted, all of these fields, including the thermal stress-energy
tensor, could of course also be written down in $3+1$-form, but as
will be made clear below, specifying these quantities in this form
will prove unnecessary for the ensuing argument.}. 

Depending crucially on whether one wishes to specify these quantities
in the so-called energy or particle reference frames, different choices
for some of the vector fields can typically be made, namely either
$\nu_{A}^{a}=-\eta^{-1}q^{a}$, $\eta^{a}=\eta^{-1}\Theta q^{a}-Q^{a}$
and $p^{a}=0$ or $\nu_{A}^{a}=0$, $\eta^{a}=\beta q^{a}-Q^{a}$
and $p^{a}=q^{a}$, where $\beta=T^{-1}\equiv(\frac{\partial\varepsilon}{\partial s})^{-1}$
is the inverse of the infamous Tolman-Ehrenfest temperature, $\eta$
is the relativistic enthalpy or 'injection energy' per particle, $\Theta$
is the so-called thermal potential of the fluid mixture (relativistic
chemical potential per particle and per temperature) and $Q^{a}$
is a second order contribution which may be interpreted as a generalized
heat flux and for which different approaches exist in the literature. 

With respect to one of these choices, the physical behavior of fluid
mixtures can be specified exactly in almost equilibrium states, in
which the expressions given below must follow the differential laws

\begin{equation}
\nabla_{a}T_{\;b}^{a}=0,\:\nabla_{a}N_{A}^{a}=0,\:\nabla_{a}S^{a}\geq0.
\end{equation}
The respective laws are the laws of local conservation of energy and
particle number and the so-called Clausius-Duhem inequality, which
is nothing more than the differential form of the entropy law of thermodynamics. 

All this is important for the subject of tidal heating for the following
reason: it makes a big difference for the form of the bulk part of
the integral relation $(12)$ whether the matter distribution under
consideration is in thermal equilibrium with its environment or not.
More precisely, it turns out that the bulk part of said relation is
always zero (or, more precisely, can be chosen to be zero) for fluid
mixtures that are in perfect thermal equilibrium, while this is not
true for corresponding matter fields outside thermal equilibrium.
For such matter distributions, as it turns out, the bulk part is always
non-zero. 

To see this, one may recall that the main equilibrium condition for
relativistic viscous fluids in extended irreversible thermodynamics
(as well as in classical theory) to be in thermal equilibrium is $\nabla_{a}S^{a}=0$;
a condition that requires the complete absence of dissipative mechanisms.
For this condition to be satisfied and for the corresponding system
to actually reach a local thermostatic equilibrium state, the heat
exchange between the fluids must come to a halt and the sum of all
thermal potentials and chemical reaction rates must approach zero.
Besides that, the thermal energy-momentum tensor and the particle
number and entropy currents of the fluid mixture must be the same
as that of an ordinary ideal fluid, meaning that $T_{\;b}^{a}=(\epsilon+p)w^{a}w_{b}+p\delta{}_{\,b}^{a}$,
$N_{A}^{a}=n_{A}\cdot w^{a}$ and $S^{a}=s\cdot w^{a}$. For this
to hold, an equilibrium equation of state of the form $\epsilon+p=T(s+\underset{A}{\sum}\Theta{}_{A}n_{A})$
must be valid, where each $\Theta_{A}$ represents a thermal potential
associated with a species $A$ of particles (characterized in terms
of the chemical potential $\mu_{A}\equiv\frac{\partial\epsilon}{\partial n_{A}}$).
Ultimately, moreover, the motion of the fluids should be rigid in
Born' s sense, which means that the orthogonal distances between the
adjacent material world lines of the two fluids must remain constant.
In effect, this means that $h_{a}^{\:c}h_{b}^{\:d}\nabla_{(c}w_{d)}=0$
must hold, which in turn means that $\nabla_{(a}w_{b)}=0$ must be
satisfied. However, from this it follows that the matter field under
consideration must, at perfect thermal equilibrium, curve spacetime
in such a way that a stationary gravitational field is generated with
a timelike Killing vector field $\xi^{a}$ that is directly proportional
to the four-velocity $w^{a}$, i.e. $\xi^{a}=\beta w^{a}$ with $\beta\equiv c\cdot N\equiv c\cdot\sqrt{-\xi_{a}\xi^{a}}$
with $c=const.$ 

And while it is also possible in principle for internal thermal equilibrium
to occur with respect to non-Killing observers, it is still reasonable
to expect that the variables $\dot{N}$, $\dot{N}^{a}$, and $\dot{h}^{ab}$
for fluid mixtures in aequilibrio are very close to zero (with respect
to the Killing time parameter, which is admittedly the most natural
choice). From this, in turn, it becomes clear that in non-equilibrium
situations the bulk part of relation $(12)$ can only be neglected
and the Thorne-Hartle model can only be in agreement with the results
of the quasilocal canonical framework of Brown and York if the $3+1$-quantities
$\varepsilon$, $p_{a}$, and $S_{ab}$ introduced in the previous
section are all exactly or at least close to zero. 

For quasilocal surfaces, which are located very far from the boundary
of the spatially finite internal gravitational field of the matter
distribution under consideration, one can safely assume that the mentioned
quantities actually become zero and thus for weak gravitational fields
the Thorne-Hartle approximation remains valid and leads to correct
results. An important reason for this is that dissipative heating
of small, weakly gravitating bodies (such as, for example, Jupiter's
satellite Io) it can be expected that only small amounts of electromagentic
radiation are emitted in the course of the tidal deformation process.
However, for gravitational interactions of more massive objects, especially
when a body is torn apart by tidal deformations or otherwise becomes
unstable during the interaction period (with the consequence that
not only the outer vacuum field of the body fluctuates, but also the
inner field, which can lead to the emission of non-negligible amounts
of electromagnetic radiation), this can no longer be expected to be
true, implying that the Hartle-Thorne approximation can no longer
be used and it becomes necessary to consider integral relation $(12)$
instead of relation $(23)$. For certain types of matter accumulations,
the structural changes in the gravitational source may even cause
mass-energy to flow out of the system at sometimes very high velocities
(think of relativistic jet flows in accreting supermassive black holes).
Such phenomena are typically expected when bodies merge with each
other; an effect that, like tidal heating, is described in various
places in the literature based on the linearized approximation of
general relativity \cite{abramowicz2013foundations,blanchet2014gravitational,burns2020neutron,chang1999pseudotensors,hamerly2011event,hussain2017deformation,schafer2018hamiltonian}. 

Now, all this has an interesting consequence: the possibility of jointly
measuring gravitational and electromagnetic radiation emitted during
violent astrophysical events such as those described above suggests
that the corrections calculated in the second part of the present
work can lead to actual observable effects. More specifically, when
considering relation (16), it becomes clear that if the Brown-York
expression is to be a physically feasible candidate for gravitational
mass-energy not only in theory, but also in reallife situations, there
should be a shift in the total intensity when electromagnetic and
gravitational radiation are emitted simultaneously from a gravitational
matter source. This effect should then be measurable. Moreover, a
comparable intensity shift should also occur when gravitoelectromagnetic
effects are measured. 

To illustrate this, the mass-energy inflow through the quasilocal
surface of a spatially and temporally bounded gravitating system caused
by gravitoelectromagnetic fields shall be calculated next. Using for
this purpose a perturbative splitting of the metric of the form $(24)$,
where the perturbation field $e_{ab}$ is used to define the trace-reversed
gravitaitonal potential $\psi_{ab}=e_{ab}-\frac{1}{2}\eta_{ab}e$,
which is subject to the conditions $\partial^{a}\psi_{ab}=0$, $\vert\psi_{00}\vert\gg\vert\psi_{jk}\vert$
and $\vert\psi_{0i}\vert\gg\vert\psi_{jk}\vert$, one finds that the
linearized Einstein's equations reduce to the form

\begin{equation}
\square\psi_{ab}=16\pi T_{ab}.
\end{equation}
Here, it can be assumed for simplicity's sake that $T_{jk}=0$ for
$j,k=1,2,3$ and therefore $S_{ab}$ applies. Moreover, it can be
assumed that $s_{ab}=0$.

By defining next the fields $\psi_{00}(x)\equiv2\phi(x)$ and $\psi_{0k}(x)\equiv2A_{k}(x)$,
one finds that the linearized field equations read

\begin{equation}
\square\phi=16\pi\varepsilon,\;\square A_{k}=16\pi p_{k},
\end{equation}
where $k=1,2,3$ shall apply in the present context. The form of the
solutions of these Poisson equations is

\begin{equation}
\phi(t,\vec{x})=\int\frac{\phi(t-\vert\vec{x}-\vec{y}\vert,\vec{y})}{\vert\vec{x}-\vec{y}\vert}d^{4}y
\end{equation}
 and

\begin{equation}
\vec{A}(t,\vec{x})=\int\frac{\vec{A}(t-\vert\vec{x}-\vec{y}\vert,\vec{y})}{\vert\vec{x}-\vec{y}\vert}d^{4}y.
\end{equation}
The resulting line element of the metric reads
\begin{equation}
ds^{2}=-\left(1-2\phi\right)dt^{2}-4A_{k}dtdx^{k}+\left(1+2\phi\right)\delta_{jk}dx^{j}dx^{k}
\end{equation}
in Cartesian coordinates. The solutions obtained can be used to define
gravielectric and gravimagnetic field strengths $\mathcal{\vec{E}}:=\frac{1}{2}\vec{\nabla}\phi$
and $\mathcal{\vec{B}}:=\vec{\nabla}\times\vec{A}$, which in turn
can be used to set up the gravitoelectromagnetic pendants to the Maxwell
equations and the Lorentz force law, i.e. the following set of relations
\begin{align}
\vec{\nabla}\mathcal{\vec{E}} & =4\pi\varepsilon\\
\vec{\nabla}\mathcal{\vec{B}} & =0\\
\vec{\nabla}\times\mathcal{\vec{E}} & =-\partial_{t}\mathcal{\vec{B}}\\
\vec{\nabla}\times\mathcal{\vec{B}} & =-4\pi\vec{p}+\partial_{t}\mathcal{\vec{E}}
\end{align}
and

\begin{equation}
\vec{F}=m\left(\mathcal{\vec{E}}+4\vec{v}\times\mathcal{\vec{B}}\right),
\end{equation}
which have to be met for the sake of consistency. 

By considering the fact that $N=\sqrt{g^{tt}}=\sqrt{1+2\phi}\approx1+\phi$,
one finds by neglecting higher order terms that relation $(12)$ takes
the form

\begin{equation}
\frac{dE}{dt}=\underset{\Sigma_{t}}{\int}(\dot{\phi}\varepsilon-p_{k}\dot{A}^{k})\omega_{h}-\underset{\Omega_{t}}{\int}(\lambda\dot{\phi}e-j_{k}\dot{\mathcal{A}}^{k}-\mathfrak{X})\omega_{q},
\end{equation}
where $\varepsilon=\frac{1}{8\pi}\left(\mathcal{\vec{E}}^{2}+\mathcal{\vec{B}}^{2}\right)$
and $p_{k}=\frac{1}{\pi}\epsilon_{kij}\mathcal{E}_{i}\mathcal{B}_{j}$
are the GEM energy current density and the GEM poynting vector and
$\mathcal{A}^{b}=q_{\;c}^{b}A^{c}$ is the vector potential on the
quasilocal boundary surface. The crucial observation in this context
is that the second corrective part $\mathfrak{X}_{2}$ of the integrand
$\mathfrak{X}$ is different from zero. It reads
\begin{equation}
\mathfrak{X}_{2}=-\lambda\eta\left(1+2\phi\right)\varepsilon+\lambda^{2}\eta\left(1+2\phi\right)p_{k}v^{k}+\lambda\eta\left(1+\phi\right)p_{k}\mathcal{A}^{k}-\left(1+2\phi\right)p_{k}u^{k}.
\end{equation}
Imagining now that the two-surface $\Omega_{t}$ is a sphere at infinity,
it becomes clear that the corresponding integral expression will also
be different form zero, i.e $\underset{\Omega_{t}\equiv S_{\infty}}{\int}\mathfrak{X}_{2}\omega_{q}\neq0$. 

From this, however, it now follows that the quasilocal treatment of
gravitoelectromagnetic effects requires consideration of the corrections
resulting from the time derivative of the Brown-York mass given in
the second section of this paper. Similar corrections should also
arise, as mentioned earlier, when gravitational and electromagnetic
radiation from a strongly gravitating matter source are measured cooperatively,
since also in such a case a non-vanishing bulk-to-boundary energy
inflow term should arise, and hence corrections to Einstein's quadrupole
formula. The reason for this is that in both cases Einstein's field
equations are satisfied for a non-vanishing matter source, which is
why the correction terms resulting from the time derivative of the
total quasilocal gravitational Hamiltonian cannot be neglected in
either case. 

From a practical point of view, the calculated corrections should
manifest themselves in the form of a shift in the intensity of the
measured radiation. It is therefore to be expected that the presence
of a corresponding detectable intensity shift of incoming gravitational
and electromagnetic radiation could provide a practical test of the
utility and feasibility of the Brown-York expression as a reasonable
candidate for the mass-energy of the gravitational field.

\section*{Conclusion }

In the present work, previous results by Booth and Creighton on quasilocal
tidal heating were extended in that the mass-energy transfer through
the quasilocal surface was calculated in a different manner, namely
by varying the total gravitational Hamiltonian (bulk plus boundary
parts) and not only the boundary part. As it turned out, all the results
obtained are fully compatible with those of Booth's and Creighton's
work if it is assumed that the vacuum field equations of the theory
are satisfied (as the authors actually did in their work). If this
is not the case, however, as was shown, correction terms corrections
resulting from the time derivative of the Brown-York mass must be
taken into account, one of which specifies how mass flows from the
bulk into the boundary. To demonstrate the feasibility of the results
obtained, their applicability to the geometric setting considered
in Booth's and Creighton's work on quasilocal tidal heating was demonstrated
in the third section of the work and a comparatively simple geometric
example was treated, namely, the mass-energy inflow through the quasilocal
surface of a spatially and temporally bounded gravitating system caused
by gravitoelectromagnetic fields. This simple example was used to
show that the matter content of a spatially and temporally bounded
gravitating system can change with time such that mass-energy flows
through the quasilocal surface and thereby escapes from the system.
From this insight, it was concluded that the obtained results can
be used to describe tidal deformation and tidal heating effects in
more challenging physical situations, such as in the description of
accretion phenomena or merger processes in relativistic $N$-body
systems.
\begin{description}
\item [{Acknowledgements:}]~
\end{description}
I want to greatly thank Ivan Booth for reading through and suggesting
improvements to the first draft of the manuscript. I would also like
to thank Friedrich Kupka for sharing his many insights into the phenomenology
of astrophysics. 

\bibliographystyle{plain}
\addcontentsline{toc}{section}{\refname}\bibliography{0C__Arbeiten_Papers_litquas}

\end{document}